# Presence in VR experiences – an empirical cost-benefit-analysis

René Peinl
Institute of Information Systems
Hof University of Applied Sciences
rene.peinl@iisys.de

Tobias Wirth
Institute of Information Systems
Hof University of Applied Sciences
tobias.wirth2@hof-university.de

**Abstract**: Virtual reality (VR) is on the edge of getting a mainstream platform for gaming, education and product design. The feeling of being present in the virtual world is influenced by many factors and even more intriguing a single negative influence can destroy the illusion that was created with a lot of effort by other measures. Therefore, it is crucial to have a balance between the influencing factors, know the importance of the factors and have a good estimation of how much effort it takes to bring each factor to a certain level of fidelity. This paper collects influencing factors discussed in literature, analyses the immersion of current off-the-shelf VR-solutions and presents results from an empirical study on efforts and benefits from certain aspects influencing presence in VR experiences. It turns out, that sometimes delivering high fidelity is easier to achieve than medium fidelity and for other aspects it is worthwhile investing more effort to achieve higher fidelity to improve presence a lot.

## Introduction

Since the availability of Oculus Rift to the public in 2016, the hype around virtual reality (VR), that was already there in the 1990ies, was revived and now seems to be here to stay. The quality of head mounted displays (HMDs) together with the low latency of sensors and enough processing power in processors and graphic cards allow for immersive experiences at comparably low cost that were not possible 25 years ago (Anthes et al. 2016). Since then, an abundance of VR companies has emerged and a number of HMDs have been made available for professionals and consumers alike. However, if you look at scientific literature, studies about presence and immersion in VR are still often conducted with specialized hardware instead of off-the-shelf devices. Additionally, although many factors influencing presence were studied, there is no analysis on the effort needed to incorporate these factors into a VR experience at a given level of fidelity. It is, e.g. clear, that visual quality positively effects presence both regarding the display of the HMD and the rendered contents, but how does effort put into render quality relate to perceived presence stemming from visual quality? Is it a linear relation, are there plateaus or does the cost-benefit function converge against an upper limit that is far below being "indistinguishable from reality"? In this study, the effects of multiple factors are measured empirically and compared to the effort needed to implement them. To do that, we carefully chose a hardware setup that has a good cost-immersion-ratio and conducted several experiments to measure the factors with the highest impact and find a good compromise between effort and perceived presence.

## Related work

Presence can be seen as the sensation of 'being there' in the virtual environment (Slater 1999). It is defined in literature as the psychological state where a user is feeling lost or immersed in the mediated environment, the degree to which he or she feels physically "present" in a virtual environment (Schubert et al. 2001). More recent interpretations make a distinction between the illusion of being in a place ('place illusion'), and experiencing events as if they were real ('plausibility') (Slater and Sanchez-Vives 2014). The former relies on head (and ideally eye) tracking to enable perception of the environment using the body in a natural way (bending down, looking around, looking past, listening by turning the head towards the source). The latter is the extent to which the events within the virtual environment are perceived as really happening (ibid) and is said to be influenced by three factors: "(a) the extent to which events in the environment refer specifically to the participant (b) the extent to which there are events that respond to the actions of the person (e.g., the participant smiles at a virtual human that smiles back) and (c) the overall credibility of the environment in comparison to expectations" (Slater and Sanchez-Vives 2014). Where presence describes the subjective feeling, immersion is the objective characteristic of the technology, referring to the extent to which the computer displays are capable of delivering an inclusive, extensive, surrounding and vivid illusion of reality to the senses of a human participant (Slater & Wilbur, 1997).

(McMahan et al. 2012) studied display and interaction fidelity of a VR first person shooter in a CAVE environment. They found that low display fidelity and low interaction fidelity together led to quickest completion time of the task, whereas "mixed fidelity" conditions (low/high or high/low) led to highest completion times. They concluded that both display and interaction fidelity had significant positive impact on presence, engagement, and usability. (Marks et al. 2014) use an Oculus Rift DK1 to study immersion in VR environments. They use the Unity game engine to visualize engineering data. However, judging visual quality based on the photos in the paper, the visual quality was much lower than the one in our bath scenario. Their workflow from the CAD software to Unity was similar to ours. The body ownership illusion is another important factor(Slater and Sanchez-Vives 2014). Experiments show that different virtual bodies even influence behavior of the user in the virtual world (ibid). (Lorenz et al. 2015) found that using body tracking with a Kinect is leading to significantly higher presence values than with a Wii controller. (Simeone et al. 2015) analyze the effect of getting haptic feedback from substitutional objects that are tracked in the real world to be accordingly represented at the same position in the virtual world. They found that substitutional objects do not have to be identical to virtual ones to enhance presence. It is sufficient if share some common characteristics like size and shape. Others like material and weight can vary up to a certain degree. (North and North 2016) report about presence in traditional and immersive virtual environments. Although this was already two years after the availability of Oculus Rift DK2, they were using a non-standard HMD with unknown resolution for the traditional VR environment and a CAVE-like room with projections on the wall as immersive environment. They found the immersive environment to enhance presence compared to the HMD. (Steed et al. 2016) found evidence that a self-avatar had a positive effect on presence and embodiment. In a different experiment with a singer inviting the participant to tap along it turned out to have a negative effect on embodiment compared to the singer not directly addressing the user. However, the results must be considered with caution, since data was collected "in the wild" and not under controlled laboratory settings. (Regenbrecht et al. 2017) report about an experiment with intentionally coarse graphics in order to decrease hardware requirements for presence research. They use a VR HMD, but use it for mixed reality, where real objects are filmed with a camera and are incorporated into the virtual world in real time. They found that intentionally low visual quality but providing a coherent experience across the whole scene with full body tracking allowed for high presence values. (Seibert and Shafer 2018) analyze the impact of a HMD on spatial presence compared to a regular computer monitor. They find that the HMD increases perceived presence. However, higher presence is not a value in itself, but leads to higher liking of the virtual place visited and therefore also a stronger intention to visit the place physically (Tussyadiah et al. 2018).

## Choice of hard- and software

Regarding hardware, we know from literature that factors effecting immersion and therefore presence in VR are visual quality, latency of reaction on user activities and the degree of freedom. Looking at off-the-shelf VR headsets we find three distinct categories (Anthes et al. 2016).

Smartphone-powered headsets are very low cost (~100€) if you already own a compatible smartphone, but provide only three degrees of freedom (3DoF). Samsung GearVR and Google Daydream are examples of this category. We do not consider "simple casings" like Google Cardboard (ibid.). PC-based headsets do have 6 DoF, but are mostly comparatively expensive (500-1500€) and are usually restricted by cables connecting them to the PC (wired), which effects mobility and perceived freedom. However, they are able to provide enough processing power for high fidelity rendering of contents, which is limited for Smartphone-powered headsets. Oculus Rift and HTC Vive are the most sold devices in this category. The last category, which is currently emerging is stand-alone devices. That means that they do not need external trackers like the HTC lighthouse system and do also not need a PC for rendering contents, but still provide 6 DoF. Oculus Quest is an example of this new device category. It is the only one that has both a 6DoF HMD and 6DoF hand controllers, which are crucial for certain types of VR experiences. This is a major improvement compared to the stand-alone systems described in (Anthes et al. 2016). Regarding display quality, there is no principle distinction between the categories of devices. You have lower quality on some PC-based HMDs like Oculus Rift than on GearVR with a Galaxy S9 smartphone, but higher quality on newly available PC-based HMDs like Pimax 8k or Varjo VR1. It has to be noted that visual quality of HMDs is not only affected by the resolution, but also by display technology since the most important aspect is the absence of the screen door effect (SDE). Samsung Odyssey+ for example achieves a lower SDE than HTC Vive due to a special light distribution foil despite having a similar resolution. SDE stems from the black areas between pixels that get visible with the large magnification of the HMD's lenses. Finally, the field of view (FoV) is important for immersion. It is around 100° for most HMDs on the market ranging from 90° to 110°. However, recently the Pimax 8k became available which offers 200° FoV at the cost of some distortions at the border of the visible field. Regarding latency, there is no great difference between the headsets, so the factor is negligible for the choice of hardware.

Due to the unavailability of the Pimax 8k during the time of the study (December 2018), we chose the HTC Vive Pro with two trackers for the lighthouse system. It was powered by a AMD Ryzen 2700X CPU, 32 GB of RAM and an Nvidia GeForce 1080 Ti GPU, so that the overall hardware invest was around 3.000€. We did not use the wireless adapter which could enhance the perceived freedom of moving. We also didn't use different HMDs as originally planned. To exclude factors we could not control, like different displays with marginally different FoV, SDE and contrast, we used the Vive Pro for all experiments including the one with 3 DoF, instead of using the Oculus Go HMD, which we had available as well. We also excluded the Lenovo Explorer which would have allowed for a larger virtual room due to its inside out tracking, which was not relevant for our experiments.

Regarding the software needed for creating the VR experience, there are two main tools needed. The 3D modelling software and the VR engine. For modelling, we preferred the open source software Blender due to its free availability over the more capable professional tools like Maya or 3D Studio max. For the VR engine, the choice is mainly between Unity and the Unreal engine, which both combine a powerful game engine for 3D rendering with a development environment with a good graphical editor. A detailed comparison of those is beyond the scope of this paper and since features are very similar, it might be at the end a subjective personal choice. (Schlueter et al. 2017) state that Unreal has superior photorealistic rendering capabilities over Unity3D, but chose Unity due to its free availability and abundance of learning resources. We chose Unity 2018.3 together with the SteamVR plugin due to existing knowledge about it. (Hilfert and König 2016) showed that the Unreal Engine is working well, too.

## Study setup

Two different VR scenarios were created, both being in the context of interior design or architecture. Commercial usages we consider are virtual exhibitions of baths, kitchens, and furniture in general as well as visualization of the results of bath makeovers or interior design. Finally, virtual visits of houses and apartments for rental or buying are possible applications. Therefore, the first scenario was a bath. We took 360° photos of a bath exhibition in Hof, Germany and recreated one of the baths as a 3D model. The photos were taken with an Insta360 Pro camera in 8k 3D and then stitched with Insta Stitcher and exported in RAW format. Exposure was enhanced using Adobe Camera Raw.

In the first experiment we compared presence with photos in full resolution (8k 3D = 7680x8640), half resolution (4k 3D = 3840x4320) as well as both resolutions in 2D. In the second experiment, we compared the photos with the 3D model. We did not put additional effort into making the 3D model look identical to the photo, but did aim for a high similarity at the least possible effort (see figures 1 and 2). We chose freely available 3D models[1] as a basis and assembled them into a room created by ourselves. The goal was to find out whether perceived visual fidelity of the contents were higher in 360° photos than in our 3D model and how that affects presence.

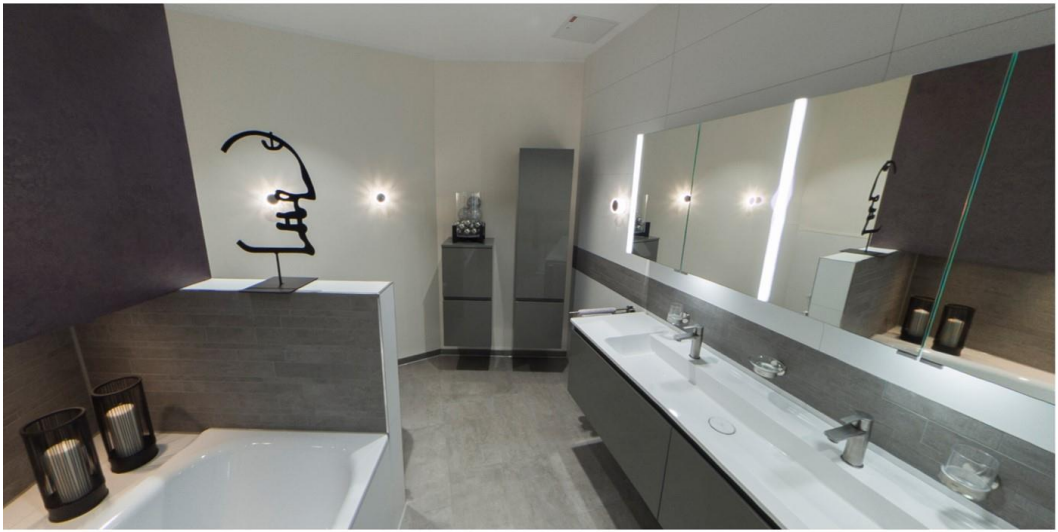

*Figure 1: 360° photo of a bath that was used in the study*

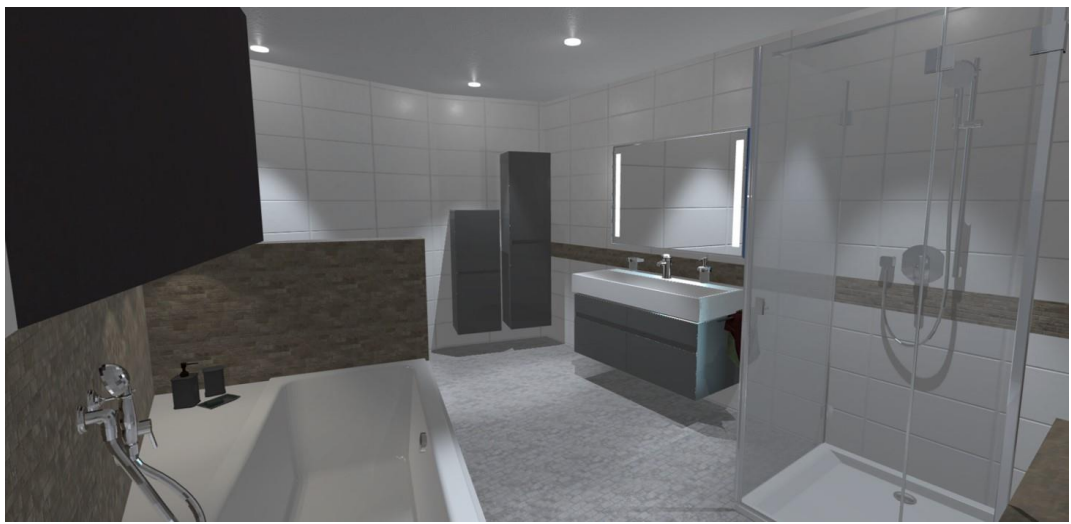

*Figure 2: 3D model of a bath, similar to the 360° photo in figure 1*

---

[1] https://free3dbase.com

The second scenario was a living room, which served as a basis to study the effects of different levels of interaction on presence. The 3D models used there were much simpler and not photorealistic at all compared to the bath scenario. The task for participants was to look for books that were distributed in the room and count those with the title "Virtual Reality".

In the first experiment, the participants could only move through the room on foot or by using the VR controllers to teleport. The second experiment allowed participants to open drawers and doors of cabinets, as well as pick-up books to better look at them. However, the interaction by intention was made very simplistic like in most computer games. Participants did only have to find the right spot with the controller, which was indicated by putting a yellow frame around the object of interaction and then press a button on the controller. A prerecorded animation was then played that showed how the door or drawer opened automatically and completely. In contrast to that, the third experiment allow for more natural interaction. The button on the controller was used for grabbing, but the movement needed to be induced by moving the arm and pulling the object. Once released the object was behaving physically correct maintaining its impulse for some time until being stopped by gravity and air resistance or friction. Additionally, books could be put everywhere around the room or thrown and also showed effects from gravity, whereas in experiment two they "magically" appeared back in their original location, once the controller button was released.

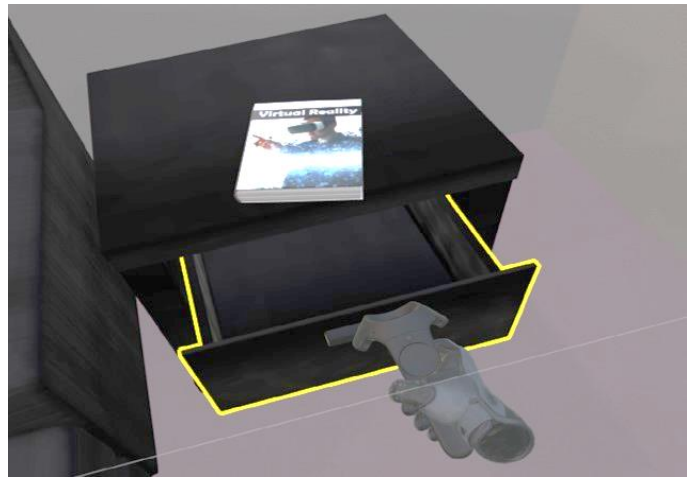

*Figure 3: virtual hand with controller opening a drawer highlighted with a yellow frame during interaction*

28 people participated in the study. They were between 18 and 52 years old (mean 30 years, sd 9.42) and were all acquired at the university (staff and students). They did not receive any payment or other benefits. There were 20 male and 8 female participants. Eight participants had at least intermediate VR experiences. Four of those did own a VR headset. The study lasted for about 30 minutes, although each experiment took only a few minutes. After each experiment, participants were asked to fill out a questionnaire to state their personal impressions on presence in various aspects. The questionnaire was mainly built with questions stemming from the presence questionnaire by (Witmer and Singer 1998) with a few own additions. We did a pretest with three people to eliminate mistakes. The questionnaire consisted of 13 questions for the bath and 19 questions for the living room scenario. They were summed up into the categories involvement, visual fidelity, adaption/immersion and interface quality (see table 1) as described in (Witmer et al. 2005). The scores per category were calculated as a sum of single question scores, divided by the number of questions. The single questions were judged on a 7-point Lickert scale from 1 being the lowest presence to 7 representing the highest presence score.

*Table 1: aggregation of questions to presence aspects*

|  | Questions bath | Questions living room |
|---|---|---|
| **Involvement** | 6, 7, 8, 12 | 3, 4, 5, 6, 7, 8, 10, 14 |
| **Visual fidelity** | 10, 11 | 12, 13 |
| **Adaption / immersion** | 9, 13 | 9,11,15,16,17,19 |
| **Interface quality** | - | 18 |

# Results

A reliability analysis of the different questionnaires resulted in a Cronbach α of 0.83 for $I_1$, 0.88 for $I_2$ and 0.84 for $I_3$. Since all of these values are well above 0.7, we can assume consistency of the answers. The same applies to the bath scenario with Cronbachs α of 0.91 for $B_1$, 0.84 for $B_2$ and 0.72 for $B_3$. The data did not follow a normal distribution and therefore we used the Wilcoxon-Mann-Whitney test (U test) to test the significance of hypothesis. The resulting p values from the U test are considered significant if they are below 5% and considered highly significant if they are below 1%.

The first hypothesis tested was

$H_1$: *presence values of 360° photos ($B_1$) are higher than those of the respective 3D model in the bath scenario, if we artificially limited the degrees of freedom of the 3D model to three ($B_2$).*

The assumption was that the visual quality of the photo is higher than that of our textured 3D model, although we put some effort in enhancing the quality of it to get near photorealism. Leveling the freedom to 3 DoF should make the photos more compelling than the 3D model. However, it turned out that $H_1$ has to be rejected. There were slight differences in the presence scores, but they are not significant and they are even slightly higher for the 3D model.

$H_2$: *presence of the 3D model limited to 3 DoF ($B_2$) will be lower than that of the same model with 6 DoF ($B_3$).*

In accordance with literature, 6 DoF makes a big difference and therefore presence values increased by nearly two, leading to a highly significant deviation (p = 0.5%). Therefore, $H_2$ was accepted.

*Table 2: results from bath scenario (SD = standard deviation)*

|  | $B_1$ mean | $B_1$ SD | $B_2$ mean | $B_2$ SD | $B_3$ mean | $B_3$ SD |
|---|---|---|---|---|---|---|
| **Involvement** | 4.259 | 0.286 | 4.490 | 0.534 | 6.241 | 0.103 |
| **Visual Fidelity** | 2.910 | 0.227 | 2.942 | 0.299 | 5.893 | 0.657 |
| **Adaption/Immersion** | 5.179 | 1.566 | 5.135 | 1.713 | 6.536 | 0.354 |
| **Presence-Score** | **4.152** | 1.068 | **4.264** | 1.140 | **6.228** | 0.378 |

The next hypothesis dealt with visual quality regarding resolution of the contents, while keeping resolution of the display constant.

$H_3$: *a higher image resolution will increase presence for 360° images.*

With p-values of 0.63% and 1.00% for 2D and 3D images respectively, $H_3$ can be accepted. An image resolution of 8k lead to significantly higher presence scores, compared to 4k.

$H_4$: *stereoscopic images (3D) will result in higher presence values for 360° photos than monoscopic.*

Again, the hypothesis can be accepted. Results were significant for both 8k images (p = 0.033%) as well as 4k images (p = 0.003%). However, some participants also reported problems with nausea that increased for 3D contents.

Table 3: visual fidelity of 360° photos in different resolutions

|  | 4k 2D | 4k 3D | 8k 2D | 8k 3D |
|---|---|---|---|---|
| mean | 3.000 | 4.679 | 4.000 | 5.429 |
| SD | 1.440 | 1.249 | 1.440 | 1.136 |

For the living room scenario, the assumption was that presence would increase with possibility and realism of interaction.

**H$_5$**: *presence values in the experiment with simplistic interaction ($I_2$) will be higher than in the experiment with no interaction ($I_1$).*

Although presence values did indeed increase for $I_2$, the difference was not significant (p = 37.02%). Therefore, H$_5$ has to be rejected.

**H$_6$**: *presence values in the experiment with realistic interaction ($I_3$) will be higher than in the experiment with simplistic interaction ($I_2$).*

Again, presence values for $I_3$ did increase compared to $I_2$. With a p-value of 2.6% the difference is significant, but not highly significant. H$_6$ was therefore accepted.

Table 4: results of the living room scenario (SD = standard deviation)

|  | $I_1$ mean | $I_1$ SD | $I_2$ mean | $I_2$ SD | $I_3$ mean | $I_3$ SD |
|---|---|---|---|---|---|---|
| **Involvement** | 4.714 | 1.118 | 5.246 | 0.585 | 5.768 | 0.449 |
| **Visual Fidelity** | 4.018 | 0.278 | 5.411 | 0.177 | 6.143 | 0.050 |
| **Adaption/Immersion** | 4.887 | 1.784 | 4.994 | 1.638 | 5.435 | 1.426 |
| **Interface Quality** | 2.071 | NA | 2.214 | NA | 2.321 | NA |
| **Presence-Score** | **4.538** | 1.421 | **4.998** | 1.235 | **5. 492** | 1.202 |

# Discussion of results in relation to effort

Some of the findings were as expected, whereas others were contrary to intuition. While increasing the resolution of images as well as adding a stereoscopic effect increases presence for 360° images as expected, it turned out that despite a perceived higher photorealism of 360° images the 3D model with high visual fidelity was preferred by participants, even if we limited it to 3DoF. One explanation for this is that some participants felt that proportions in the 360° photo were not right, so it felt like the floor or ceiling were too far away. Another part of the explanation is that current computing power and game engines are able to produce a visual quality that is getting closer and closer to photorealism. This is especially true for static scenes where lights, reflections and shadows can be precomputed. Unity makes it possible to achieve good results even for rather unexperienced 3D designers like us. We did not even use the post-processing stack or the new HD render pipeline, which can further enhance visual quality. The effort is also not too high overall (see Table 5), although it is significantly higher than producing a 360° photo. The effort pays off, however, if you look at the increase in presence of 2. Although the scale is ordinal and not metric, this is a large difference. The importance of visual quality can also be seen when comparing B3 with I1. Although the comparison is not perfectly adequate since the environment differs not only in visual fidelity, it shows a clear trend. Interaction, however, is also an important aspect. Despite a lower rise in presence scores of "only" ~1.5 between $I_1$ and $I_3$ in relation to ~2.0 from $B_1$ to $B_3$, it can be stated that interaction should not be neglected. This is especially true in relation to efforts, since using the SteamVR plugin together with its PreFabs[2] makes implementing realistic interaction including physics easy. The feature "throwable" makes any 3D object immediately pickable and behaving according to physic laws (esp. gravity). In addition, friction can be simulated very easy. Both only need additional effort of a few minutes. We however found, that the simulated friction does not behave realistically in all situations. It works well for inclined planes as underground, but fails for horizontal grounds that are moved under a second object lying upon it. For drawers and cabinet doors we were using the PreFabs "linear drive" and "circular drive" from SteamVR. Again, they only need a few minutes to be added to 3D models that are rigged[3] properly. Astonishingly, the simplified interaction tested in $I_2$ is not only less effective in terms of increasing presence, but also causes more effort than the realistic interaction implemented in $I_3$. Therefore, it is a clear advice to use the latter kind of interaction.

---

[2] a Unity-specific term denoting a prebuilt bundle that can easily be added to a scene and may include 3D models, materials, textures as well as scripts that are then directly usable.

*Table 5: overview of tasks for creating VR experiences*

| Task | Learning effort | Implementation effort |
|---|---|---|
| 1) Create 360° photo | Very low (minutes) | Short setup per location plus 15 min per photo |
| 2) Create 3D model<br>    a. acquire<br>    b. create on your own | Finding right platforms<br>~1 day of learning Blender basics | ~15 min per object<br>1 hour up to days depending on complexity |
| 3) Create materials and textures<br>    a. acquire<br>    b. create on your own | Finding the right platforms<br>~1 day learning 3D photo scanning basics | 15-30 min per material<br>1-4 hours per material |
| 4) Fix 3D models or materials | ~ 4 h of learning UV mapping and normals | 10 min – 4 hours per object or material |
| 5) Assemble the 3D scene | ~1 day learning Unity basics | 2-8 hours depending on the complexity of the scene |
| 6) Set up lighting and reflections | ~4h of learning basics | 1-4 hours depending on the complexity of the scene |
| 7) Apply the post processing stack | ~ 4h of learning basics | ~1 hour experimenting with the settings to get a good result |

On the other hand, it was often harder than expected to integrate existing 3D models downloaded from the internet. In contrast to the PreFabs mentioned, 3D models in obj or fbx format often needed a considerable amount of fixing before they could be used in Unity. This applied to sizes as well as normals and materials including textures. Unity uses sizes where 1 unit should relate to 1 m in the real world. Imported 3D models often were either factor 100 or even 1000 too large, or factor 10 or 100 too small. This was however easy to fix. Normals facing in the wrong direction led to invisible surfaces. This was fixed in Blender. You can either flip the normals to the other side there, or you can let Blender recalculate them, which usually leads to the desired results. (Marks et al. 2014) report about similar issues when exporting their models from the CAD tool. The most effort went into fixing materials and textures. There seems to be no working standard that lets you smoothly transport those from a 3D modeling tool to Unity. It is unclear, whether it is a Unity problem or existent in Unreal Engine as well. Basically all materials have to be recreated. The only benefit you have is that the assignment of materials to objects or faces already exists. This problem even exists for Blender files, which are explicitly supported in Unity, so there is one less conversion. The problem is amplified by the fact that both in Blender and in Unity, there are two different render engines for normal and high quality that are not directly compatible. In Blender, you have the common renderer and Cycles, the physically based renderer. Similarly, you can create either a normal 3D project in Unity or one with the HD render pipeline. In any case, you cannot directly use materials from the other version.

---

[3] Prepared to be movable in the game

## Conclusion and outlook

Our empirical study has shown that it is possible to create VR experiences with a high level of presence with comparatively low effort. Visual quality of the content, 6 DoF and realistic interaction with virtual objects are important factors affecting presence. Current off-the-shelf VR headsets are a good basis for VR experiences, since they provide a high level of immersion. The screen door effect and the low field of view were the main concerns people had regarding the hardware. Interesting side aspects discovered via free text comments of participants or oral utterances were that the cable of the Vive Pro did bother multiple people. In future experiments we plan on using wireless transmission of signals, although they could negatively affect visual quality and latency. We also found that 6DoF can reduce nausea compared to 3DoF, whereas 3D photos tend to increase it. This is however only anecdotal evidence and needs to be confirmed in further studies.

Another aspect not studied enough in literature is breaking the feeling of presence by having virtual objects obstruct the real body of the user. Multiple people reported that they were pushed out of the presence illusion when they opened a cabinet door and it easily moved through the position where their virtual body should be. This happened although we did not use a virtual avatar, so it could not be seen directly, that the door was moving through the virtual knee. For a follow-up study we envision to use an avatar animated on the basis of human pose detection (Mehta et al. 2017) and implement collision detection between virtual objects and this avatar. This should help avoiding these kinds of irritations. However, it cannot be avoided completely since no physical resistance will hinder users to still walk through virtual objects. Maybe the avatar could be kept in front of the virtual obstacle, but it would still be hard to show the real movement of feet and legs, while doing so. However, we could not observe people walking through virtual objects in our current study.

Another interesting aspect we completely neglected for this study is audio and especially positional audio. For the next study, we plan on analyzing the effects of simple audio and spatial audio effects on presence and relate it to the respective efforts. Related work indicates that VR experiences benefit from spatial sound (Poeschl et al. 2013). In order to make results from single experiments more comparable, we envision a single scenario for all different experiments, no matter whether it is audio, visual or interaction experiments. With the availability of Pimax 8k and HP Reverb in Q2/2019, it would also be interesting to investigate the effects of different hardware with differences in screen door effect, field of view and controller quality on presence.